\documentclass[twocolumn,aps,showpacs,superscriptaddress,prb,tightenlines,amsmath,amssymb]{revtex4-1}
\usepackage{comment}
\usepackage[colorlinks, linkcolor=blue, anchorcolor=blue, citecolor=blue]{hyperref}

\usepackage{graphicx}
\usepackage{amssymb}
\usepackage{dcolumn}
\usepackage{amsmath}
\usepackage{physics}
\usepackage{mathrsfs}  
\usepackage{color}
\usepackage{bm}
\usepackage{colordvi}

\makeatletter

\newcommand{\Rmnum}[1]{\expandafter\@slowromancap\romannumeral #1@}
\makeatother

\begin{document}

\title{Soliton Pumping in the Rice–Mele Model with On-Cell Kerr Nonlinearity }
\author{Zhe Wang}
\affiliation{{Key Laboratory of Quantum Information,
University of Science and Technology of China, CAS, Hefei, Anhui
230026, People's Republic of China}\\
{Synergetic Innovation Center of Quantum Information and Quantum Physics, University of Science and Technology of China, Hefei, Anhui 230026, China}}
\author{Xi-Wang Luo}
\email{
luoxw@ustc.edu.cn}
\affiliation{{Key Laboratory of Quantum Information,
University of Science and Technology of China, CAS, Hefei, Anhui
230026, People's Republic of China}\\
{Synergetic Innovation Center of Quantum Information and Quantum Physics, University of Science and Technology of China, Hefei, Anhui 230026, China}}
\author{Bo-Ye Sun}
\email{boyesun@ytu.edu.cn}
\affiliation{YanTai University, Yantai, Shandong, 264005, People's Republic of China}
\author{Zheng-Wei Zhou}
\email{zwzhou@ustc.edu.cn}
\affiliation{{Key Laboratory of Quantum Information,
University of Science and Technology of China, CAS, Hefei, Anhui
230026, People's Republic of China}\\
{Synergetic Innovation Center of Quantum Information and Quantum Physics, University of Science and Technology of China, Hefei, Anhui 230026, China}}
\date{\today}

\begin{abstract}
We investigate the Rice–Mele model with on-cell Kerr-type nonlinearities, where the interaction depends on the total particle number within each unit cell rather than on individual sites. This interaction enables a nontrivial interplay between topology and nonlinear dynamics in soliton pumping. In the weakly interacting regime, the ground-state soliton undergoes quantized Thouless pumping. At intermediate interaction strengths, soliton creation and annihilation break adiabaticity and disrupt quantized transport. In the strong-coupling regime, the coexistence of ground- and excited-state solitons leads to negligible coupling at energy crossings, giving rise to discrete time-translation symmetry breaking (DTTSB) in the soliton dynamics. Comparison of mean-field results with exact diagonalization along closed circular pumping paths confirms both the validity of the mean-field description and the robustness of DTTSB across different pumping trajectories. Our findings reveal how interaction-induced effects can fundamentally modify topological transport and suggest that these phenomena may be explored in cold-atom, photonic, and superconducting-circuit platforms.
\end{abstract}

\maketitle

\section{Introduction}
Topological phases of matter have become a central focus in modern condensed matter physics~\cite{klitzing1980new,thouless1982quantized,Kane2005Z2Topological,haldane2015model,citro2023thouless,Hasan2010Colloquium,qixiaoliang2011Topological,Cooper2019Topological,Ozawa2019Topological}. Unlike conventional phases characterized by local order parameters, topological phases are distinguished by global invariants that are robust against local perturbations. A paradigmatic example is the Thouless pump, a one-dimensional system with adiabatically modulated parameters in which quantized transport directly reflects the topological nature of the band structure~\cite{thouless1983Quantization,experimental2018ma,luo2018Topological,Topological2018Kolodrubetz,ma2025Topological,citro2023thouless,Jurgensen2021Quantized,yang2024Nonadiabatic}.

Since its proposal, the Thouless pump has been generalized to bosonic systems and realized in a variety of synthetic dimensions and engineered platforms~\cite{jaeyooncho2008Fractional,experimental2018ma,luo2018Topological,Trautmann2024relaization,Topological2018Kolodrubetz,liaoyuwei2025Realization,Padhan2024Padhan,Ravets2025thouless,Julie2024Quantized}. More recently, nonlinear extensions have attracted attention, particularly the transport of solitons in nonlinear media~\cite{Jurgensen2021Quantized,Jurgensen2022chern,yefangwei2022nonliner,mostaan2022quantized,Jurgensen2023Quantized,Tuloup_2023,INSPEC:26606811,you2025Nonlinear,xiao2025nonlinear,tao2024nonlinearity,xuyong2025Nonlinearity}. In the weakly interacting regime, both theory and experiment~\cite{Jurgensen2022chern,mostaan2022quantized} have demonstrated that solitons still exhibit quantized displacement determined by the Chern number of the underlying band structure, with their motion closely tracking adiabatic Wannier trajectories. However, the fate of such quantized pumping in strongly interacting or structurally unconventional nonlinear systems remains largely unexplored.

To address this, we study a Rice–Mele model with an unconventional on-cell interaction. While conventional Hubbard-type interactions act independently on each lattice site, a two-site unit cell allows more general interactions that include cross terms between sublattices~\cite{tao2024nonlinearity,sone2025transition}. Of particular interest is the case where the cross-interaction equals twice the on-site term, rendering the interaction quadratic in the total particle number per unit cell. In this limit, the interaction energy depends only on the unit-cell occupation, not on its distribution between sublattices. Consequently, a soliton confined within a unit cell is unaffected by interactions, while its motion between unit cells can be strongly modified. By effectively “screening” intra-cell effects yet preserving inter-cell interaction-induced dynamics, this setup provides a minimal and analytically tractable platform to understand how strong correlations can qualitatively modify topological pumping.

In this work, we show that such interactions fundamentally change the soliton dynamics in a pump cycle. In the weakly interacting regime, the ground-state soliton performs a quantized Thouless pump. Beyond a critical interaction strength, however, the pumping protocol induces the creation of additional solitons, which annihilate the original one and destroy quantized transport. Upon further increasing the interaction strength, the annihilation is delayed, and in the strong-coupling regime, the soliton requires two pump periods to return to its initial position, realizing spontaneous discrete time-translation symmetry breaking (DTTSB). Using exact diagonalization, we further demonstrate that this phenomenon is not tied to a particular pumping protocol but persists for more general circular paths in parameter space.

Our work thus identifies the Rice–Mele model with on-cell interactions as a simple setting where topology, strong correlations, and nonequilibrium symmetry breaking meet, opening a route to controlled studies of DTTSB phases in clean, tunable systems.

\section{\label{sec:model}Model and calculation method}
\begin{figure}[h]
\includegraphics[width=0.48\textwidth]{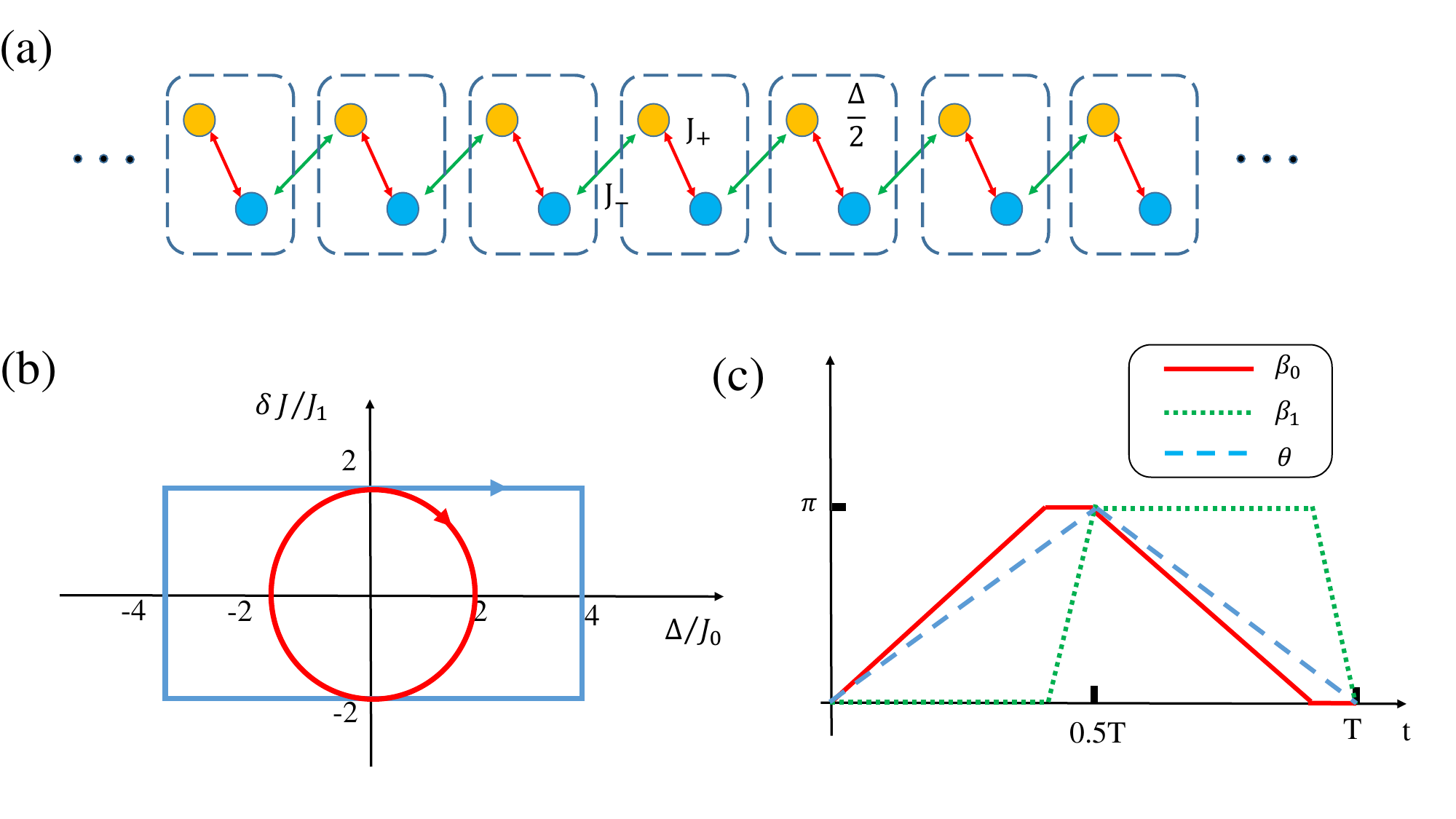}
\caption{\textbf{Rice–Mele model and pumping protocol.} (a) Rice–Mele model with two sites per unit cell, opposite on-site potentials, and unequal intra- and inter-cell hoppings. (b) Pumping trajectories in the $\delta J$–$\Delta$ plane: the blue square loop is experimentally convenient, while the red circular loop illustrates the generality of the results. (c) Time evolution of the pump-protocol parameters $\beta_0$ and $\beta_1$ used in the calculations, as well as the parameter $\theta$ for circular pumping [Eqs.~\eqref{eq:circ_Delta} and \eqref{eq:circ_J}].}\label{fig1}
\end{figure}

In this section, we introduce our model and the corresponding calculation method. The total Hamiltonian can be written as $H=H_{\mathrm{RM}}+H_{\mathrm{Kerr}}$. Here, $H_{\mathrm{RM}}$ is the 
Rice–Mele model~\cite{rice1982elementary}:
\begin{eqnarray}
    \mathrm{H}_{\mathrm{RM}}  &=&-\sum_{j}(J_{+}b_{j,2}^{\dagger}b_{j,1}+J_{-}b_{j+1,1}^{\dagger}b_{j,2}+\mathrm{H.c.})\nonumber\\
  +&&\sum_{j}\frac{\Delta}{2}(b_{j,1}^{\dagger}b_{j,1}-b_{j,2}^{\dagger}b_{j,2}),\label{eq:rm}
\end{eqnarray}
where $b_{j,\alpha}$ is the annihilation operator at the unit cell $j$ and the sublattice $\alpha=1,2$. The intra-cell tunneling amplitude $J_+$ differs from the inter-cell tunneling amplitude $J_-$, and $\Delta$ represents the sublattice onsite energy offset as shown in Fig.~\ref{fig1}(a). 

The $H_{\mathrm{Kerr}}$ term represents the interaction, which can be generally written as 
\begin{eqnarray}
H_{\mathrm{Kerr}} &=& -\kappa \sum_j [b^\dagger_{j,1}b_{j,1}(b^\dagger_{j,1}b_{j,1}-1)+2\gamma b^\dagger_{j,1}b_{j,1}b^\dagger_{j,2}b_{j,2} \nonumber\\
&+&b^\dagger_{j,2}b_{j,2}(b^\dagger_{j,2}b_{j,2}-1)],
\end{eqnarray}
where $b^\dagger_{j,1}b_{j,1}$ and $b^\dagger_{j,2}b_{j,2}$ are the particle numbers in the two sublattices of the unit cell $j$, $\kappa$ denotes the strength of the Kerr nonlinearity, and $\gamma$ tunes the strength of the cross-interaction term. If $\gamma=1$, this nonlinear interaction term becomes $(b^\dagger_{j,1}b_{j,1}+b^\dagger_{j,2}b_{j,2})(b^\dagger_{j,1}b_{j,1}+b^\dagger_{j,2}b_{j,2}-1)$, i.e., the on-cell total-number interaction, which is the focus of this work.

For the Rice–Mele model, it is well known for the existance of the topological pumping when the closed pump loop traced out by the detuning $\Delta$ and the tunneling imbalance $\delta J = |J_+| - |J_-|$ encloses the critical point at $\Delta = \delta J = 0$. Moreover, the topological properties of the pump remain invariant under any continuous deformation of this loop, as long as it does not cross the critical point\cite{thouless1983Quantization,luo2018Topological,mostaan2022quantized}. In this work, the pumping is controlled by a time variation of the detuning 
\begin{eqnarray}
J_\pm &=& J_1\left(1 \pm e^{i\beta_1(t)}\right),\label{eq:pump_Jpm}\\
\Delta(t) &=& -4J_0 \cos[\beta_0(t)], \label{eq:pump_Delta}
\end{eqnarray}
with $J_0$, $J_1$ denotes the maximum energy offset and hopping strength, respectively, and $\beta_0(t)$, $\beta_1(t)$ describes the pump protocol.

For analytical and numerical convenience, we adopt a square-shaped pumping loop in the 
$(\Delta, \delta J)$ parameter space, indicated by the blue square in Fig.~\ref{fig1}(b). Along this loop, the pump parameters $\beta_0(t)$ in Eq.~\eqref{eq:pump_Delta} and $\beta_1(t)$ in Eq.~\eqref{eq:pump_Jpm} are varied alternately: one parameter increases linearly by $\pi$ while the other is held constant, and then the roles are switched. Specifically, as shown in Fig.~\ref{fig1}(c), from $0$ to $ t_1$, $\beta_0(t)$ increases linearly by $\pi$ while $\beta_1(t)$ remains constant; from $t_1$ to $t_1+t_2$, $\beta_1(t)$ increases linearly by $\pi$ while $\beta_0(t)$ is held fixed. Repeating this alternating pattern once completes a full cycle with both parameters reaching $2\pi$, where we choose $t_1 = 9t_2$. The slow modulation of $\beta_0(t)$ and $\beta_1(t)$ ensures adiabatic evolution throughout the pumping process.

With the model, we can now calculate its time evolution. The time evolution of the annihilation operator is governed by the Heisenberg equation:
\begin{equation}
    i \dot b=[b,\mathrm{H}],
\end{equation}
where $\mathrm{H}= \mathrm{H}_{\mathrm{RM}}+ \mathrm{H}_{\mathrm{Kerr}}$ denotes the total Hamiltonian of the system. 

In our system, the photon number is sufficiently large that their behavior can be treated classically. This justifies the use of the mean-field approximation, where quantum fluctuations are negligible. By taking $\phi_{j,\alpha} = \langle b_{j,\alpha} \rangle = \langle b_{j,\alpha}^\dagger \rangle$, the dynamics reduce to a discrete nonlinear Schr\"odinger equation\cite{mostaan2022quantized,kevrekidis2009The,raghavan1999coherent}
\begin{eqnarray}
    i\dot \phi_{j,1}&=&-J_+ \phi_{j,2}-J_-\phi_{j-1,2}+[\frac{\Delta}{2}-2\kappa n_j - \kappa]\phi_{j,1} \label{eq:gp1}\\
i\dot \phi_{j,2}&=&-J_+ \phi_{j,1}-J_-\phi_{j+1,1}-[\frac{\Delta}{2}+2\kappa n_j + \kappa]\phi_{j,2}, \label{eq:gp2}
\end{eqnarray}
with $n_j=|\phi_{j,1}|^2+|\phi_{j,2}|^2$. The total particle number of the system can be expressed as $N = \sum_{j,\alpha} |\phi_{j,\alpha}|^2$. For convenience, we impose a normalization condition $\sum_{j,\alpha} |\phi_{j,\alpha}|^2 = 1$. With this normalization, the nonlinear interaction strength can be described by $g = 2\kappa N$. 

Given these nonlinear Schrödinger equations, the instantaneous eigenstates for each set of pump parameters along the loop are obtained using a self-consistent iterative method. These stationary solutions provide a reference for analyzing the system’s behavior during the pumping process. The full pumping dynamics, on the other hand, are simulated by directly integrating the time-dependent nonlinear Schrödinger equations, starting from the ground state corresponding to the initial pump parameters under the prescribed protocol~\cite{yangjianke2010Nonlinear,Tuloup_2023}.

\section{Results}

In this section, we systematically investigate the dynamical behavior of soliton pumping during the transition from weak to strong nonlinear interaction regimes, based on the nonlinear Schr\"odinger equation. Numerical simulations are performed on a system consisting of 20 unit cells, with the period of each pumping cycle lasting $T=40\pi/J_1$. The tunneling amplitude and on-site potential strength are set to $J_1=1$ and $J_0=5$, respectively.

\subsection{Weak interaction regime}
\begin{figure}[h]
  \includegraphics[width=0.48\textwidth]{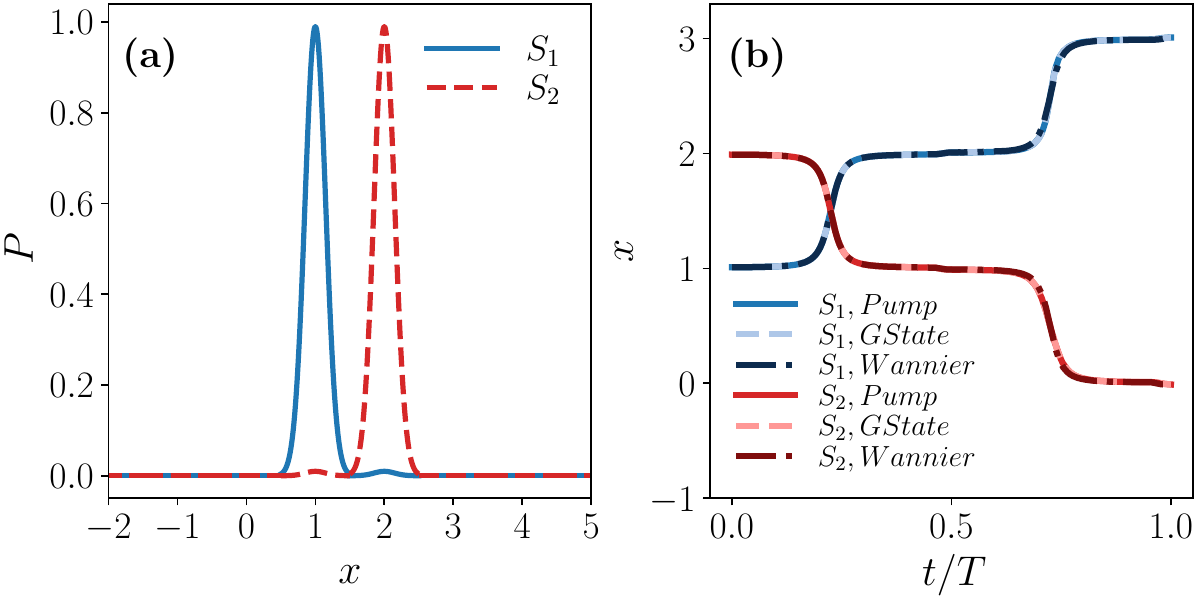}
  \caption{\textbf{Instantaneous soliton wave function distribution and its displacement over time for the case with $g=J_1$.} 
(a) Wave functions of the two-soliton solutions of Eqs.~\eqref{eq:gp1} and \eqref{eq:gp2} at $t=0$. Each lattice site is represented by a Gaussian wave packet whose width matches the grid spacing, and height is scaled to the corresponding probability amplitude for visual clarity. 
(b) Time evolution of the center of mass of the pumped solitons from simulations (solid curves, labeled 'Pump'), the instantaneous soliton solutions of Eqs.~\eqref{eq:gp1} and \eqref{eq:gp2} (dashed curves, labeled 'GState'), and the corresponding noninteracting Wannier wave function (dash-dotted curves, labeled 'Wannier'). The $y$-axis denotes the sublattice positions, where two sublattices constitute a single unit cell.
}
\label{fig2}
\end{figure}
We now investigate the system's dynamics under weak nonlinearity, e.g., $g=J_1$,  using the normalized nonlinear Schr\"odinger equation. To prepare the initial states, we first solve the equation using the self-consistent iterative method.  This yields two self-consistent soliton solutions, denoted as $S_1$ and $S_2$, whose spatial probability distributions are shown in Fig.~\ref{fig2}(a)\cite{chiao1964selftrapping,Lumer2013selflocalized}. These solutions are then individually used as initial states for real-time evolution, during which the system is subjected to a pump cycle defined by the blue square loop in Fig.~\ref{fig1}(b). The resulting center-of-mass displacements over one complete pumping cycle are plotted in Fig.~\ref{fig2}(b), labeled as "$S_1,\  Pump$" and "$S_2,\ Pump$", respectively. As expected in the weakly interacting regime, the center-of-mass shifts by one unit cell (i.e., two sublattices), reflecting quantized transport with a Chern number equal to one\cite{asboth2016ashortcourse,luo2018Topological}.

For comparison, we also compute the center-of-mass shifts of two Wannier states under the same parameters. Their trajectories during the Thouless pumping process, labeled as '$S_{1/2},\ Wannier$' and shown by the dashed-dot curves, are presented in Fig.~\ref{fig2}(b). The results show that the soliton and Wannier center-of-mass displacements are identical, consistent with the findings of Mostaan et al.~\cite{mostaan2022quantized}.

Moreover, according to the nonlinear adiabatic theorem, the soliton is expected to adiabatically follow the instantaneous eigenstate of the Hamiltonian\cite{gang2017Adiabatic,carles2012Semiclassical}. To verify this, we calculate the instantaneous eigenstates of the Hamiltonian at each time step. Their center-of-mass positions, denoted by '$S_{1/2},\ GState$' and shown by dashed curves, are also plotted in Fig.~\ref{fig2}(b). The excellent agreement between the soliton trajectory and that of the instantaneous eigenstates confirms that, in the weakly nonlinear regime, soliton transport faithfully remains adiabatic throughout the pumping cycle.

\subsection{Media interaction regime\label{sec:med}}
\begin{figure}[h]
  \includegraphics[width=0.48\textwidth]{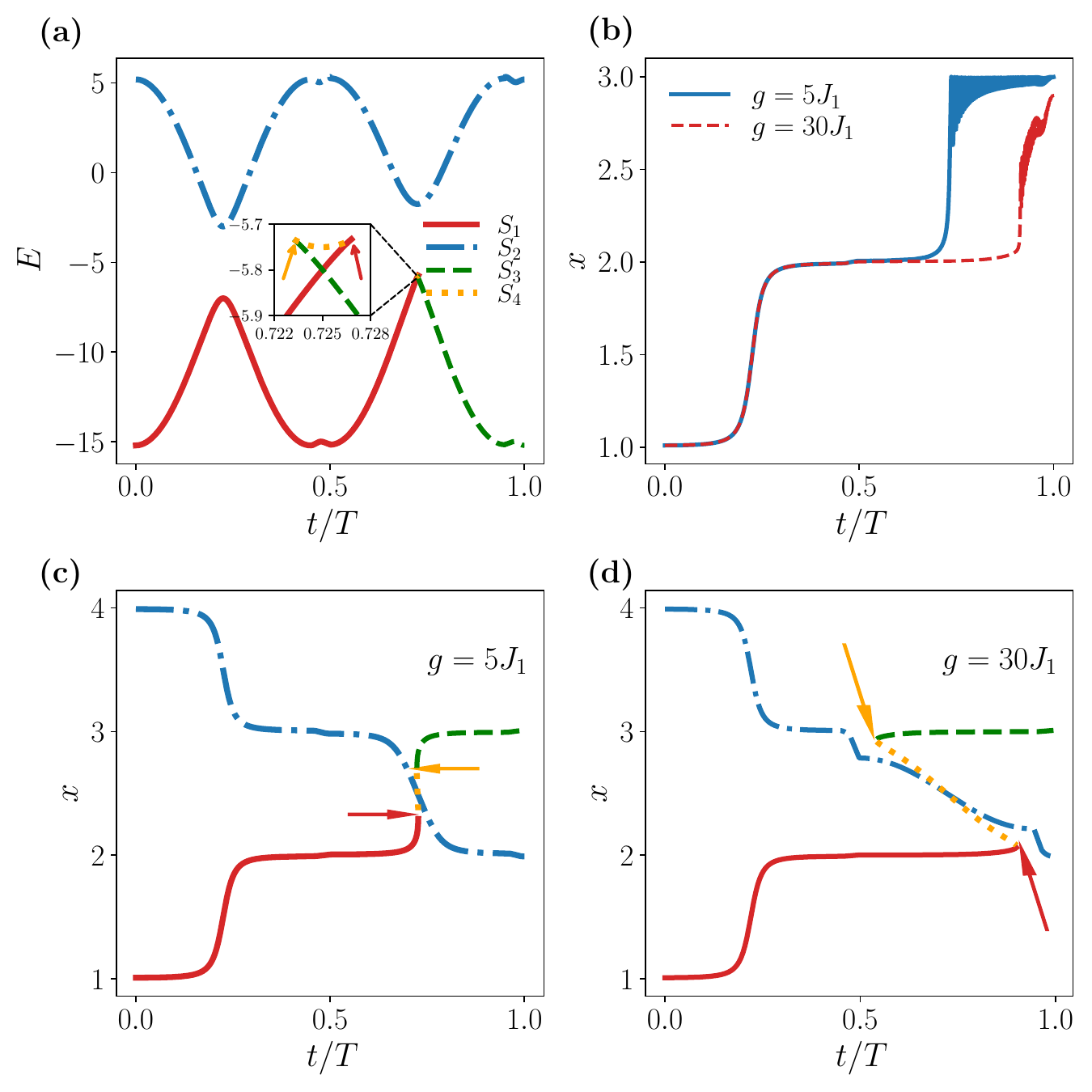}
\caption{\textbf{Soliton spectrum and trajectories at intermediate interaction strength.}
(a) Soliton energy spectrum. Inset is the magnified view near the energy crossing point, where soliton creation and annihilation occur.
(b) Time evolution of the soliton position expectation value for $g = 5J_1$ (blue) and $g = 30J_1$ (red), showing the onset of rapid oscillations.
(c) and (d) Instantaneous soliton positions from Eqs.~\eqref{eq:gp1} and \eqref{eq:gp2} for $g = 5J_1$ and $g = 30J_1$, respectively. Discontinuity of the instantaneous solution associated with $S_1$ indicates a soliton transition between different localized modes, producing the rapid oscillations in (b). Yellow arrows mark soliton creation events and red arrows mark annihilation events.}
\label{fig3}
\end{figure}
As the interaction strength increases, the system’s instantaneous energy spectrum undergoes significant changes. At the symmetric point $\Delta=0$ and $J_+=0$, i.e., the $\delta J=-2J_1$ and $\Delta=0$ point in Fig.~\ref{fig1}(b), where the Hamiltonian possesses a $\mathbb{Z}_2$ symmetry under sublattice exchange, a critical transition occurs when the interaction strength $g$ exceeds $4J_1$ (see App.~\ref{app:Z2}). This transition is marked by the emergence of a cross structure in the instantaneous spectrum, as shown in the inset of Fig.~\ref{fig3}(a). As further illustrated in this figure, two additional soliton branches, $S_3$ (green dashed) and $S_4$ (yellow dotted), spontaneously appear before this critical point (marked by a yellow arrow). Beyond the symmetric point, the $S_3$ branch becomes the new ground state of the instantaneous Hamiltonian, whereas the $S_4$ branch eventually annihilates with the $S_1$ branch (marked by the red arrow). Since the pumping process initially follows the $S_1$ branch, which serves as the ground state at the start, the annihilation of the $S_1$ branch signifies the breakdown of adiabatic pumping. 

It is important to note that the $S_3$ and $S_4$ solitons are not obtained using the self-consistent iterative method described in Sec.~\ref{sec:model}, as they correspond to saddle points of the energy functional rather than stable minima. To identify these states, we instead determine the extrema of the energy functional. In the third stage of the pumping process, where the intracell hopping is effectively switched off, the soliton becomes well localized on a single pair of neighboring lattice sites. Under this condition, the putative saddle-point solitons can be reliably obtained by extremizing the corresponding energy functional.

Since the system is initialized in the ground-state soliton $S_1$ at $t = 0$, the disappearance of this state induces a non-adiabatic transition to the lower-energy branch $S_3$. This transition manifests as late-time oscillations in the pumped soliton dynamics, as shown in Fig.~\ref{fig3}(b), consistent with recent observations by Xiao et al.~\cite{Tuloup_2023,you2025Nonlinear,xiao2025nonlinear}. To further confirm this behavior, we track the time evolution of the center-of-mass position of the instantaneous soliton solutions, as shown in Fig.~\ref{fig3}(c). The results clearly show that the final position of the pumped soliton tends to $x = 3$, coinciding with the location of the instantaneous $S_3$ soliton (green dashed curve).

Another aspect to note is that the annihilation event is progressively delayed as the interaction strength increases, as shown in the $g = 30J_1$ case in Fig.~\ref{fig3}(d), where the annihilation occurs (red arrow) much closer to $T$ compared to Fig.~\ref{fig3}(c). When the interaction strength further increases and the annihilation time eventually exceeds the pump period $T$, the $S_1$ soliton remains stable throughout the cycle, marking the onset of the strong interaction regime.

\subsection{Strong interaction regime\label{sec:strong}}
\begin{figure}[h]
  \includegraphics[width=0.48\textwidth]{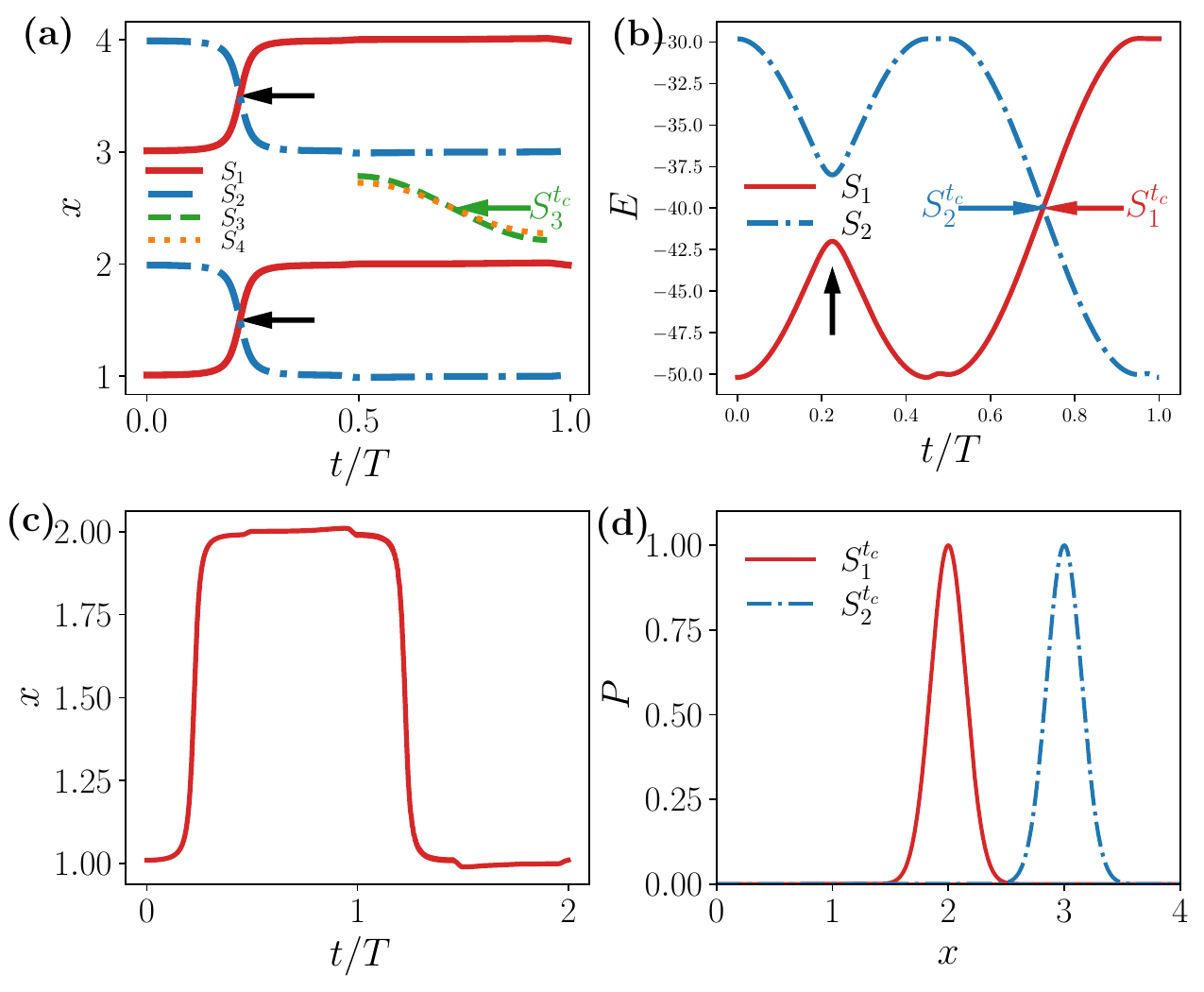}
\caption{\textbf{Dynamical soliton trajectory and instantaneous properties for $g=40J_1$.}
(a) Center-of-mass trajectory of the instantaneous soliton solutions. The pumped soliton primarily follows the $S_1$ branch, with a single crossing point with the $S_2$ branch, marked by the black arrow. Owing to translational symmetry, $S_1$ and $S_2$ are identical to their unit-cell-shifted counterparts, which are also shown.
(b) Time evolution of the energies of $S_1$ and $S_2$. At the trajectory crossing point (black arrow), the two branches remain gapped, ensuring adiabatic evolution. An additional level crossing between $S_1$ and $S_2$ occurs at $t = t_c$ (marked by blue and red arrows), where the corresponding solitons are denoted as $S_1^{t_c}$ and $S_2^{t_c}$. 
(c) Time evolution of the center-of-mass position of the pumped soliton, exhibiting period-doubling behavior, where the soliton completes a full cycle only after $2T$.
(d) Wavefunction distributions at $t_c$. The spatial separation between $S^{t_c}_1$ and $S^{t_c}_2$ further preserves adiabaticity in the pumped soliton evolution. Each site’s distribution is depicted as Gaussian wave packets [as in Fig.~\ref{fig2}(a)] for visual clarity.}
\label{fig4}
\end{figure}
Numerical simulations show that this regime sets in once the interaction strength exceeds a critical value (approximately $g>31.3J_1$ for the parameters considered here). Beyond this threshold, the $S_1$ soliton persists throughout the pumping cycle without instability. For clarity, we illustrate this regime in Fig.~\ref{fig4}(a) using the instantaneous soliton solutions at $g=40J_1$. Owing to the translational symmetry of soliton solutions, i.e., a soliton shifted by one unit cell remains a valid solution, we show both the original $S_{1}$, $S_2$ and their unit-cell-shifted counterpart in Fig.~\ref{fig4}(a) to better visualize the relative positions of instantaneous solitons. Moreover, although two additional solitons, $S_3$ and $S_4$, still emerge spontaneously in this case, they no longer annihilate with $S_1$, allowing the pumped soliton to stably follow the trajectory of the $S_1$ branch.

However, near the end of one pump cycle, $S_1$  becomes a high-energy soliton as shown by the red curve in Fig.~\ref{fig4}(b), requiring an additional cycle to return to its original ground-state position. This behavior is reflected in the center-of-mass trajectory shown in Fig.~\ref{fig4}(c), where the pumped soliton returns to its initial position only after two full cycles. This indicates spontaneous DTTSB.

Notably, while the trajectories of $S_1$ and $S_2$  overlap in real space, as marked by the black arrow in Fig.~\ref{fig4}(a), their energies differ significantly, as shown in Fig.~\ref{fig4}(b). Hence, their spatial crossing does not affect the adiabatic evolution. On the other hand, at their energy crossing point $t=t_c$ marked by the blue and red arrows in Fig.~\ref{fig4}(b), the wavefunctions of the two solitons, denoted by $S_1^{t_c}$ and $S_2^{t_c}$, are highly separated in space, as depicted by the blue and red curves in Fig.~\ref{fig4}(d). This combined protection from energy-level separation and spatial wavefunction decoupling ensures the adiabaticity of the evolution and is essential for the emergence of DTTSB in our system.

To provide an intuitive picture of this mechanism, we consider a minimal schematic scenario. In the strong-interaction regime, the on-cell attractive interaction strongly suppresses the intercell motion. This allows us to focus on the soliton motion between the two sublattice sites within a single unit cell, described by the effective Hamiltonian
\begin{eqnarray}
\mathrm{H}&=&-(J_{+}b_{2}^{\dagger}b_{1}+\mathrm{H.c.})+\frac{\Delta}{2}(b_{1}^{\dagger}b_{1}-b_{2}^{\dagger}b_{2})\nonumber\\
&-&\kappa\bigl(b^\dagger_{1}b_{1}+b^\dagger_{2}b_{2}\bigr)\bigl(b^\dagger_{1}b_{1}+b^\dagger_{2}b_{2}-1\bigr).
\end{eqnarray}
Here the nonlinear term depends only on the total occupation $N_{\mathrm{cell}}=b^\dagger_{1}b_{1}+b^\dagger_{2}b_{2}$ of the unit cell. Consequently, once the soliton is localized inside a cell, the interaction energy becomes insensitive to its distribution between the two sublattice sites, and the intra-cell configuration is primarily governed by the on-site energy imbalance $\Delta$. This effective description captures the essential ingredients of the pumping dynamics at the single-cell level, where the interplay between $\Delta$ and $J_+$ controls the soliton motion.

Within this simplified picture, one pump cycle can be viewed as follows:
(i) adiabatically turning on the intracell tunneling ($J_+$ increases from $0$ to $2J_1$);
(ii) reversing the on-site energy imbalance $\Delta$ to transfer the soliton between the sublattice sites;
(iii) adiabatically turning off the tunneling; and
(iv) restoring the original on-site energies.
Although the Hamiltonian returns to its initial form after one complete cycle, the soliton does not necessarily do so. Once the tunneling is switched off ($J_+=0$) after the soliton has moved [i.e., after step (iii)], it becomes pinned to a sublattice site. The subsequent step (iv), which restores the on-site energies while the tunneling remains suppressed, prevents the soliton from retracing its original path within the same cycle. Only in the next pump cycle, when the tunneling is re-enabled, can the soliton complete its round trip. This mechanism naturally accounts for the observed period doubling in the center-of-mass trajectory and highlights how the on-cell interaction gives rise to spontaneous DTTSB.

\subsection{General pumping loop and exact calculation}
\begin{figure}[h]
  \includegraphics[width=0.48\textwidth]{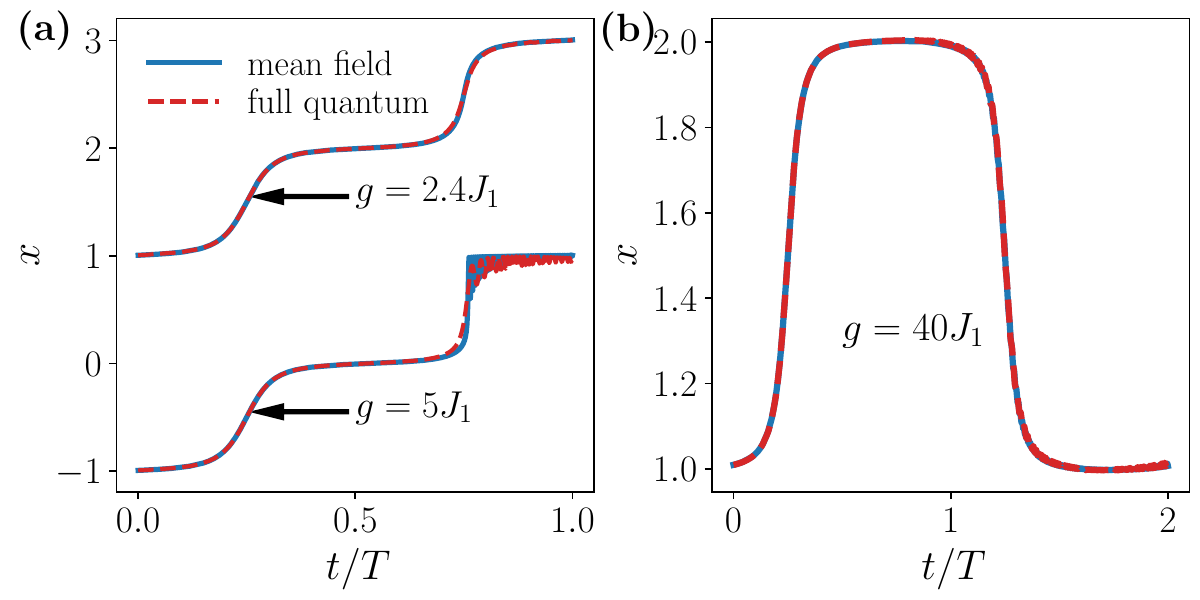}
\caption{\textbf{Center-of-mass position of pumped solitons for different interaction strengths.} (a) Blue solid lines: mean-field results for $g=2.4J_1$ and $5J_1$ under a circular pumping loop, matching the corresponding exact quantum simulations (red dashed lines). (b) Same setup as in (a) but for $g=40J_1$, showing similar agreement between mean-field and exact results.}
\label{fig5}
\end{figure}
To facilitate theoretical analysis, our study so far has adopted a square-shaped pump trajectory [Fig.~\ref{fig1}(b), blue curve] and employed a mean-field framework. However, in realistic experiments, a circular pump protocol is more commonly implemented, and fluctuations beyond mean-field may also play a role. To verify the robustness of our results, we first compute the soliton trajectory under the circular pump using the mean-field approach. The circular trajectory we consider [Fig.~\ref{fig1}(b), red circle] is parametrized by
\begin{eqnarray}
\Delta &=& 2 J_0 \cos[\theta(t)],\label{eq:circ_Delta}\\
J_\pm &=& J_1 \left\{1 \pm \sin[\theta(t)]\right\},\label{eq:circ_J}
\end{eqnarray}
where $J_0=5$, $J_1=1$, and  $\theta(t)$ increases slowly and adiabatically from $0$ to $2\pi$.

Similar to the square protocol, we observe quantized Thouless pumping in the weak interaction regime [Fig.~\ref{fig5}(a), blue solid curve with $g=2.4J_1$], oscillatory behavior at intermediate interaction strength [Fig.~\ref{fig5}(a), blue solid curve with $g=5J_1$], and spontaneous DTTSB in the strong interaction regime [Fig.~\ref{fig5}(b), blue solid curve].

To further assess the validity of the mean-field approximation in our calculation, we perform exact numerical simulations for a finite system of 5 unit cells with 6 particles. Specifically, we obtain the initial ground states via imaginary time evolution and track the soliton position by solving the time-dependent Schr\"odinger equation with the exact Hamiltonian $H_{\mathrm{RM}}+H_{\mathrm{Kerr}}$ during the pump. The resulting pump trajectories remain in agreement with the mean-field predictions, as illustrated in Fig.~\ref{fig5}. Moreover, the fidelity between the mean-field and exact ground states exceeds 99.9\%, reinforcing the reliability of our mean-field analysis. These results are consistent with previous work~\cite{buonsante2005attractive,kanamoto2006critical}, which demonstrated that in the strong-attraction, self-trapped soliton regime, mean-field results agree well with exact calculations.

It is worth noting that in the fully quantum regime, a weak coupling can emerge between the two soliton states at the level-crossing point, i.e., $S_1^{t_c}$ and $S_2^{t_c}$. The coupling strength decreases exponentially with increasing particle number~\cite{Karkuszewski2002Mean,wubiao2006Commutability,Graefe2007Semiclassical}. Consequently, the system realizes a prethermal discrete time crystal with an exponentially long lifetime in the fully quantum case, whereas it becomes a genuine discrete time crystal within the mean-field approximation.

\section{Conclusion and discussion}
In summary, we have investigated the Rice–Mele model with on-cell Kerr-type nonlinearities, where the interaction energy depends on the total particle number within each unit cell rather than on individual sites as in the conventional Bose–Hubbard framework. This distinctive form of interaction enables a rich interplay between topology and nonlinear dynamics in the context of soliton pumping.
For weak interactions, the ground-state soliton faithfully follows the adiabatic pump cycle, realizing quantized transport as expected from Thouless pumping. Once the interaction strength exceeds $g>4J_1$, however, the pump evolution drives the creation of additional solitons, which subsequently annihilate the original ground-state soliton. This process breaks adiabaticity and disrupts the quantized pump. By further increasing the interaction strength, the onset of such annihilation is delayed, and in the strong-coupling regime, the newly created and original solitons coexist without annihilation within a single pump period. In this limit, the coupling between ground- and excited-state solitons at energy crossings becomes negligible, while their separation at position crossings grows significantly. The resulting dynamics display DTTSB in the pumped soliton trajectories.

We have also confirmed the validity of the mean-field picture by comparing it with exact diagonalization of small systems along a circular pumping path, which not only reproduces the soliton trajectories across different interaction regimes but also demonstrates that DTTSB persists beyond the specific protocol, holding for more general closed paths in parameter space. Taken together, these results highlight how interaction-induced effects can fundamentally alter topological pumping and give rise to emergent DTTSB in nonlinear lattice systems. 

Experimentally, the Rice–Mele model can be realized by atoms in optical superlattices~\cite{lohse2016thouless}, photons in coupled waveguides/cavities, electric/superconducting circuits~\cite{tao2025emulating,athanasiou2024thouless}, and even by synthetic dimensions constructed by photonic and atomic internal degrees of freedom~\cite{luo2018Topological,zhuchenxi2022Correlated}.  The sublattices can be encoded in atomic spins or photonic polarizations that share the same spatial site, which would give rise to on-cell interactions~\cite{tao2024nonlinearity,sone2025transition}.
The on-cell interaction may also be engineered in electric circuits~\cite{zhou2022topological}.
Therefore, the phenomena explored here could be within experimental reach, opening a path to controlled studies of interaction-driven breakdown of quantized transport and time-translation symmetry breaking.

\begin{acknowledgments}
The authors thank Xian-Hao Wei for his valuable suggestions. This work is supported by the National Natural Science Foundation of China (Grants No. 12204399, No. 12474366, and No. 11974334) and the Quantum Science and Technology-National Science and Technology Major Project (Grant No. 2021ZD0301200). X.-W. Luo also acknowledges support from the USTC start-up funding.
\end{acknowledgments}

\appendix

\section{\label{app:Z2}Spontaneous $\mathbb{Z}_2$ symmetry breaking at the sublattice-symmetric and $J_+=0$ point}

When the on-site potential vanishes, i.e., $\Delta=0$, the system exhibits a $\mathbb{Z}_2$ symmetry under sublattice exchange. Denoting the sublattice-exchange operation as $\hat{U}$, any eigenstate $\phi$ of the Hamiltonian, i.e., $H\phi = E\phi$, satisfies $H\hat{U}\phi = E\hat{U}\phi$. If $\hat{U}\phi$ is not proportional to $\phi$, the ground state at this symmetric point is necessarily degenerate.

When the intra-cell tunneling vanishes ($J_+=0$) while the inter-cell tunneling $J_-=2J_1$ remains finite, the ground-state wavefunction localizes predominantly on the two neighboring sites connected by $J_-$. Let $\phi_2$ and $\phi_3$ denote the magnitude of the wavefunction on these two sites and take $\theta$ being their relative phase, the corresponding energy function, restricted to these two sites, is then
\begin{eqnarray}
    h=\langle H \rangle&=&-2J_-\phi_2 \phi_3 \cos\theta-\kappa N (\phi_2^4+\phi_3^4)\nonumber\\
    &=& 2\kappa N \phi_2^2 \phi_3^2-2J_- \phi_2 \phi_3\cos\theta-\kappa N. \label{eq:Z2}
\end{eqnarray}
with normalization $\phi_2^2 + \phi_3^2 = 1$. Under sublattice exchange, the wavefunction transforms as $\phi=(\phi_2, \phi_3) \to \hat{U}\phi=(\phi_3, \phi_2)$. We consider the soliton solution related to the ground state where $\cos\theta$ should be one to minimize $h$.

The energy depends quadratically on the product $\phi_2 \phi_3$, which is bounded by $0 \le \phi_2 \phi_3 \le 1/2$ as a consequence of the normalization condition $\phi_2^2 + \phi_3^2 = 1$. Minimizing the energy yields
\begin{equation}
    \phi_2\phi_3= min\left \{ \frac 1 2,\frac{J_-}{2\kappa N} \right \}.
\end{equation}
From this value and the normalization condition, $\phi_2$ and $\phi_3$ can be determined. When $\phi_2 \phi_3 \ne 1/2$ (i.e., for $\kappa N > J_-$ or equivalently $g>4J_1$), the ground state is no longer invariant under $\hat{U}$, signaling spontaneous breaking of the $\mathbb{Z}_2$ symmetry and the emergence of a degenerate ground-state manifold. The originally symmetric ground state has now become a saddle point, resulting in the unstable structure shown in Fig.~\ref{fig3}(c).

\section{\label{Different_gamma}Stability of DTTSB for $\gamma \neq 1$}

\begin{figure}[t]
  \includegraphics[width=0.48\textwidth]{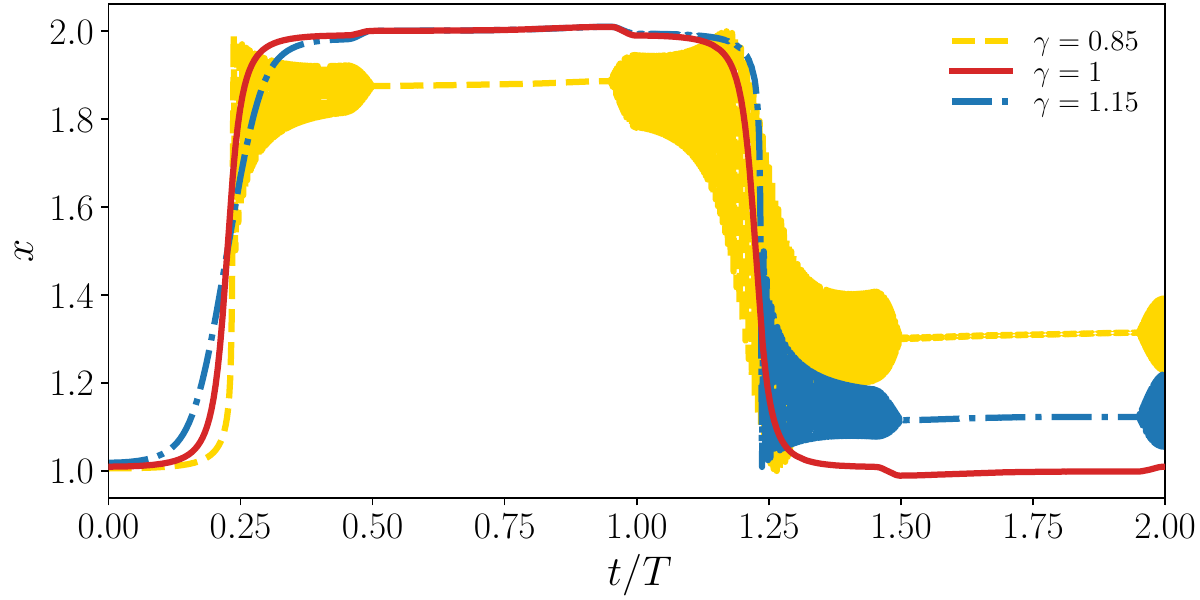}
\caption{\textbf{Breakdown of DTTSB away from $\gamma=1$.}  
Time evolution of soliton trajectories for $\gamma=0.85$ (yellow), $\gamma=1$ (red), and $\gamma=1.15$ (blue), with $g=40J_1$. While $\gamma=1$ exhibits robust period-doubled motion, even modest deviations destabilize the trajectories and destroy DTTSB. This demonstrates that DTTSB persists only within a narrow window around $\gamma=1$ (approximately $0.9 \lesssim \gamma \lesssim 1$ with $g=40J_1$), highlighting the special role of the $\gamma=1$ point.}
\label{figapp}
\end{figure}

In the main text, we focused on the case $\gamma=1$, where the interaction reduces to a purely on-cell form and allows for a simple intuitive picture of DTTSB. This picture, however, does not directly generalize to other values of $\gamma$. To test the robustness of DTTSB, we examine soliton transport for different $\gamma$ with $g=40J_1$. As illustrated in Fig.~\ref{figapp}, for $\gamma=0.85$ (yellow) and $\gamma=1.15$ (blue), the soliton trajectories become unstable and the period doubling disappears. These examples demonstrate that DTTSB crucially relies on the special interaction structure near $\gamma=1$. At the same time, our numerical checks indicate that small deviations from $\gamma=1$ do not immediately destroy DTTSB: the effect remains robust within a narrow window, approximately $0.9 \lesssim \gamma \lesssim 1.1$ at $g=40J_1$, highlighting the special role of the $\gamma=1$ point.

A qualitative understanding of this stability window follows from a simple two-mode approximation. At the special point where the on-site potential vanishes $(\Delta=0)$ and intercell tunneling is suppressed $(J_-=0)$, strong attraction confines the wave function to the two sites of a unit cell, which can be described by amplitudes $(\phi_1,\phi_2)$. Within a mean-field treatment the energy functional is
\begin{equation}
h = \langle H \rangle = -2 J_+ \phi_1 \phi_2 \cos \theta - \kappa \bigl(\phi_1^4 + 2\gamma \phi_1^2\phi_2^2 + \phi_2^4 \bigr),
\end{equation}
subject to $\phi_1^2+\phi_2^2=1$, where $g=2\kappa$ and $\theta$ is the relative phase.

In the region $\gamma<1$, we consider the ground-state soliton, and energy minimization requires $\cos\theta=1$, reducing the energy to
\begin{equation}
h=-2J_+\phi_1\phi_2 - g(\gamma-1)\phi_1^2\phi_2^2.
\end{equation}
As a function of $z=\phi_1\phi_2\in[0,1/2]$, $h$ is quadratic and opens upward, with symmetry axis
\begin{equation}
z_c=\frac{-J_+}{g(\gamma-1)}.
\end{equation}
Under the normalization condition, $\phi_1=1/\sqrt{2}$ (i.e., $z=1/2$) is always an extremum. If $z_c$ lies outside the physical domain, this remains the sole extremum, corresponding to the ground-state soliton during pumping. When $z_c$ enters the domain, however, an additional extremum emerges, signaling the birth of a new soliton that annihilates the original one, as in Fig.~\ref{fig3}, thereby destroying DTTSB.

The critical values $\gamma^c$ at which this occurs are given by
\begin{equation}
g(\gamma^c-1)= 2J_+.
\end{equation}
For $g=40$ and $J_+=-2$, this yields $\gamma>0.9$ for $\gamma<1$. A similar analysis for $\gamma>1$ gives the complementary bound $\gamma<1.1$. Together these results predict a stability window $\gamma\in(0.9,1.1)$, consistent with our numerical observation that DTTSB persists only near $\gamma=1$.

\section{Effect of the on-site potential on the interaction threshold for DTTSB}
\begin{figure}[h]
  \includegraphics[width=0.48\textwidth]{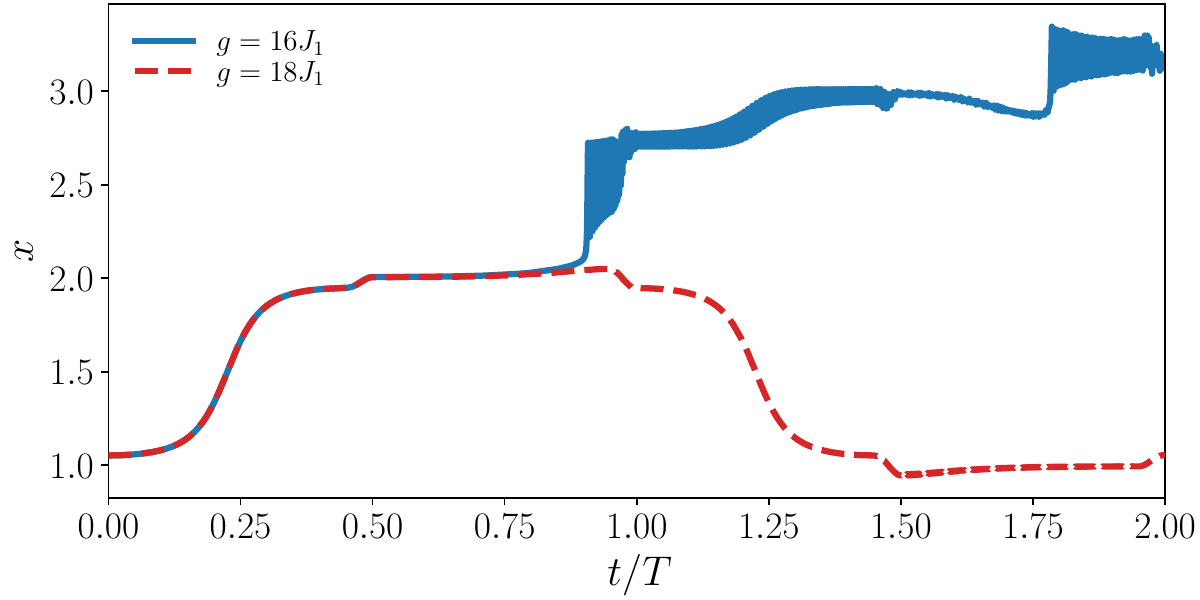}
\caption{\textbf{Effect of the on-site potential on the interaction threshold for DTTSB.}  DTTSB emerges at a lower interaction strength when the on-site energy is reduced. For $J_0=2J_1$, DTTSB occurs already at $g=18J_1$ (red curve), whereas for $g=16J_1$ (blue curve) the system remains in the intermediate-interaction regime without DTTSB. Compared with the larger on-site potential analyzed in the main text (Sec.~\ref{sec:strong}), the threshold is thus significantly lowered.}
\label{figapp2}
\end{figure}
The emergence of DTTSB requires sufficiently strong attractive interactions to confine the soliton within a single unit cell, preventing it from leaking into neighboring cells. During the third stage of pumping, the soliton is transferred to the next unit cell by tuning the on-site potential. Thus, whether DTTSB occurs depends on the competition between the on-site potential and the nonlinear interaction strength. A smaller on-site potential lowers the interaction threshold, since a weaker attraction is already sufficient to localize the soliton within a cell.

To illustrate this effect, we compare two representative cases with reduced on-site energy $J_0 = 2J_1$. As shown in Fig.~\ref{figapp2}, DTTSB already emerges at $g=18J_1$ (red curve), whereas at $g=16J_1$ (blue curve) the system remains in the intermediate-interaction regime discussed in Sec.~\ref{sec:med} of the main text, without exhibiting DTTSB. Compared with the larger on-site potential considered earlier, the threshold for DTTSB is thus significantly lowered.

 \section{BdG Spectrum and Stability Analysis}

To analyze the stability of soliton solutions, we expressing the field operators as $\psi_{j,\alpha} = \langle \psi_{j,\alpha} \rangle + \delta \psi_{j,\alpha}$ with $\langle \psi_{j,\alpha} \rangle$ being the mean field solution of the Gross-Pitaevskii equations, and the operator $\delta \psi$ describes the corresponding fluctuation field. Then the BdG Hamiltonian is obtained as follows:\cite{pitaevskii2003bose} 
\begin{equation}
    \omega \begin{bmatrix}
    \delta\psi \\ \delta\psi^\dagger
    \end{bmatrix} 
    = \mathscr{L}_{BdG} 
    \begin{bmatrix}
    \delta \psi \\ \delta \psi^\dagger
    \end{bmatrix} = \begin{bmatrix}
    A & B \\ -B^* & -A^*
    \end{bmatrix}\begin{bmatrix}
    \delta\psi \\ \delta\psi^\dagger
    \end{bmatrix} .
\end{equation}
Where 
\begin{align}
A_{j,\alpha;j',\beta} &=
H^{RM}_{j,\alpha;j',\beta} 
- \kappa\!\left(2N_j^{(0)} - 1\right)\!\delta_{jj'}\delta_{\alpha\beta}
 \\ & \quad -2\kappa\,\delta_{jj'}\,\psi^{(0)}_{j,\alpha}\psi^{(0)*}_{j,\beta}
- \mu\,\delta_{jj'}\delta_{\alpha\beta},
\label{eq:A_matrix}\\[6pt]
B_{j,\alpha;j',\beta} &=
- 2\kappa\,\delta_{jj'}\,\psi^{(0)}_{j,\alpha}\psi^{(0)}_{j,\beta},
\label{eq:B_matrix}
\end{align}
and $H^{RM}$ is the Hamiltonian of the linear Rice-Mele model.

\begin{figure}[h]  
    \includegraphics[width=0.46\textwidth]{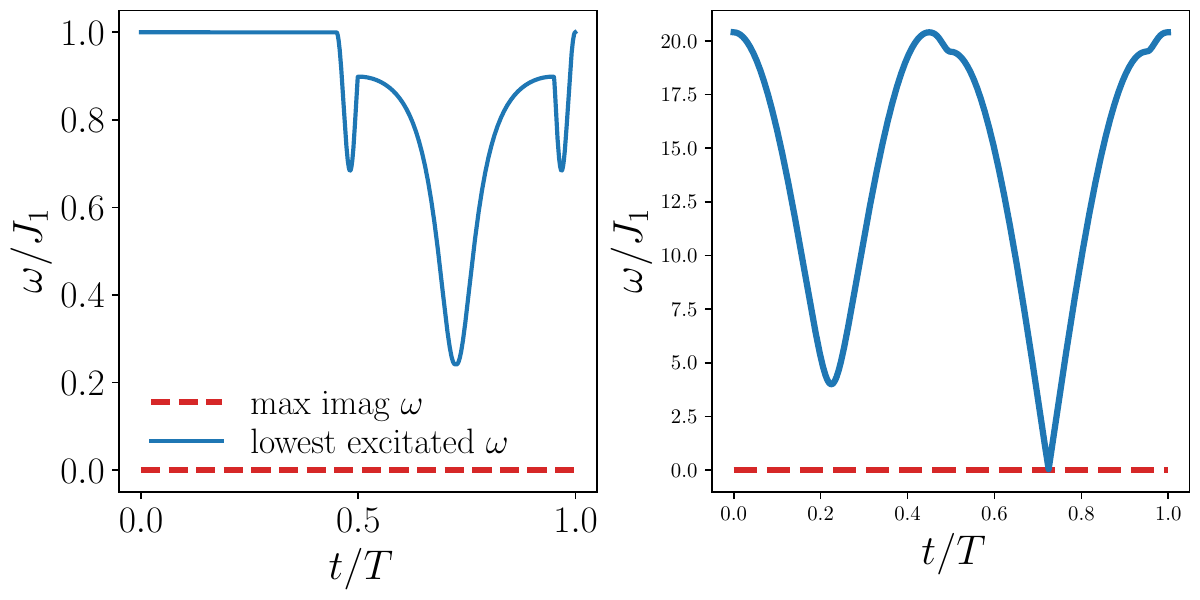}  
    \caption{\textbf{Maximum imaginary part and BdG spectrum of the lowest excitation state.}
The red dashed curves represent the maximum imaginary part of the BdG spectrum, while the blue solid curves show the BdG spectrum of the lowest excitation state. (a) For $g=J_1$, the excitation spectrum remains gapped throughout the pumping cycle, ensuring adiabatic evolution. (b) For $g=40J_1$, a single gapless point appears in the excitation spectrum; nevertheless, as discussed in the text, adiabatic evolution can still be maintained. In both cases, the imaginary parts of the BdG spectra remain zero, confirming the dynamical stability of the soliton solutions.}
    \label{fig:FigApp4}  
\end{figure}
For weak nonlinearity regime, the imaginary parts of the BdG excitation spectrum are all zero throughout the pumping process, and all excitations are protected by an energy gap as shown in Fig.~\ref{fig:FigApp4}(a). Thus, as long as the pumping speed is not excessively fast, the pumping process under weak nonlinear interactions is stable.

\begin{figure}[h]  
    \includegraphics[width=0.46\textwidth]{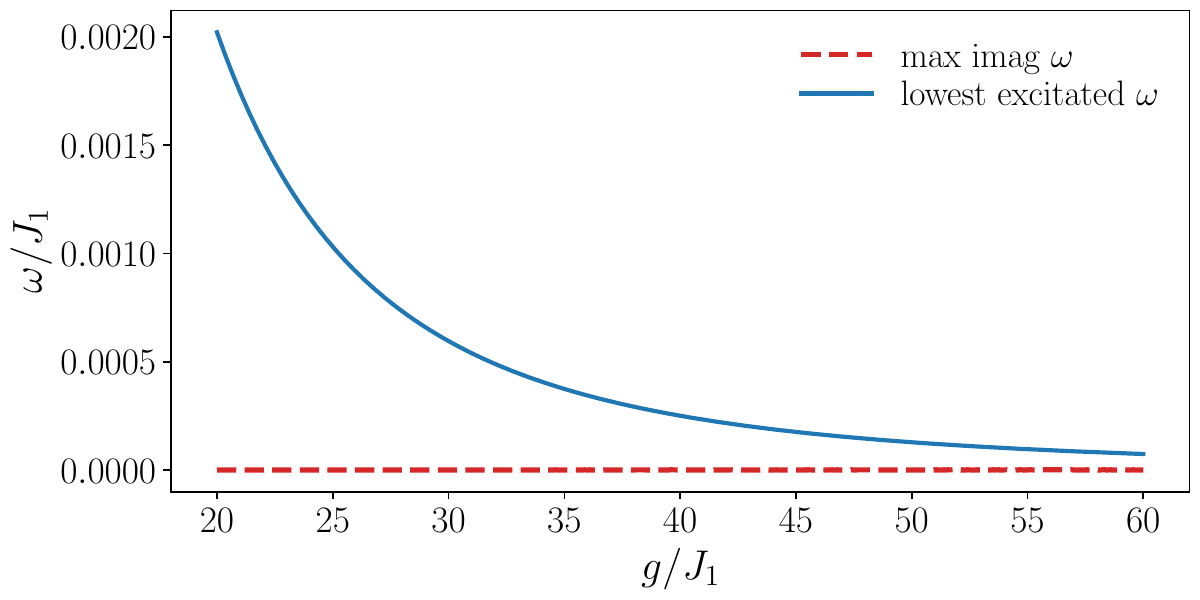}  
    \caption{\textbf{Maximum imaginary part and BdG spectrum of the lowest excitation state in $t_c$ point with different $g$.}
At the $t_c$ point, the largest imaginary part of the BdG spectrum (red curve) remains zero for all nonlinear strengths, $g$, while the smallest BdG excitation (blue curve) decreases with increasing $g$. The observed dependence of the gap on the parameter $g$ confirms its physical origin and effectively rules out the possibility of a numerical error.}
    \label{fig:FigApp4_1}  
\end{figure}

For strong nonlinearity, similar to the weak-interaction case, the imaginary parts of the BdG excitation spectrum remain zero, confirming the dynamical stability of the soliton. As shown in Fig.~\ref{fig:FigApp4}(b), although a sizable energy gap is maintained for most of the pumping process, the BdG energy gap becomes extremely small (numerically, $\omega \sim 10^{-4}$) near the $t_c$ point [see Fig.~\ref{fig4}]. To verify that this tiny gap is not a numerical artifact, we further investigated the BdG spectrum at this point for various nonlinear strengths $g$, as shown in Fig.~\ref{fig:FigApp4_1}. The results show that the imaginary parts of the BdG spectrum remain zero for all $g$, while the energy of the lowest BdG excitation gradually decreases with increasing interaction strength. Nevertheless, a finite mini-gap persists in the range of $\sim 10^{-3}$ to $10^{-4}$, confirming that it represents a genuine physical feature rather than a computational inaccuracy.

Although this mini-gap becomes extremely small near $t = t_c$, its presence still protects the soliton from nonadiabatic transitions. This robustness can be understood from the fact that the coupling between the ground state and the quasi-zero-energy excitation is proportional to the time derivative of the intra-cell tunneling, ${\partial_t J_+}$. In the pumping loop considered here, this derivative vanishes exactly at $t = t_c$, i.e., $\partial_t J_+ = 0$, effectively suppressing transitions between the two states despite the near-degeneracy. Consequently, during real-time evolution, the soliton remains protected by this finite mini-gap, ensuring adiabatic evolution throughout the pumping cycle.

\bibliography{ref}

\end{document}